\documentclass[a4paper]{article}
\pdfoutput=1
\usepackage{jheppub}

\usepackage{amsmath}

\usepackage{amssymb}
\usepackage{graphicx,color}

\usepackage{amsmath,amsthm}
\usepackage{txfonts}

\usepackage{mathrsfs}

\usepackage{bm}

\newcommand{\Ref}[1]{(\ref{#1})}
\newcommand{\Z}{\mathbb{Z}}
\newcommand{\R}{\mathbb{R}}
\newcommand{\C}{\mathbb{C}}

\newcommand{\half}{\frac{1}{2}}
\newcommand{\ab}{{ab}}

\newcommand{\bA}{{\bar{A}}}

\newcommand{\Slc}{\mathrm{SL}(2,\mathbb{C})}
\newcommand{\PSlc}{\mathrm{PSL}(2,\mathbb{C})}
\newcommand{\slc}{\fs\fl_2\mathbb{C}}

\newcommand{\Su}{\mathrm{SU}(2)}

\def\be{\begin{eqnarray}}
\def\ee{\end{eqnarray}}

\newcommand{\ci}{\mathcal I}
\newcommand{\cj}{\mathcal J}

\newcommand{\cl}{\mathcal L}
\newcommand{\cm}{\mathcal M}

\newcommand{\co}{\mathcal O}
\newcommand{\cp}{\mathcal P}

\newcommand{\cs}{\mathcal S}

\newcommand{\sa}{\mathscr{A}}
\newcommand{\sm}{\mathscr{M}}

\newcommand{\ff}{\mathfrak{f}}

\newcommand{\fl}{\mathfrak{l}}

\newcommand{\fs}{\mathfrak{s}}  \newcommand{\Fs}{\mathfrak{S}}

  \newcommand{\Fx}{\mathfrak{X}}

\renewcommand{\a}{\alpha}

\newcommand{\g}{\gamma}
\newcommand{\G}{\Gamma}

\newcommand{\eps}{\varepsilon}

\newcommand{\sig}{\sigma}
\newcommand{\Sig}{\Sigma}
\renewcommand{\l}{\lambda}
\renewcommand{\L }{\Lambda}
\renewcommand{\o}{\omega}
\renewcommand{\O}{\Omega}
\renewcommand{\t}{\tau}

\newcommand{\rmd}{\mathrm d}

\newcommand{\lt}{\left}
\newcommand{\rt}{\right}

\newcommand{\rag}{\right\rangle}

\newcommand{\tr}{\mathrm{tr}}

\newcommand{\Ar}{\mathbf{a}}

\newcommand{\sgn}{\mathrm{sgn}}

\newcommand{\bfmu}{\pmb{\mu}}
\newcommand{\bfnu}{\pmb{\nu}}
\newcommand{\bfm}{\mathbf{m}}
\newcommand{\bfn}{\mathbf{n}}

\newcommand{\sz}{\mathscr{Z}}

\title{Loop-Quantum-Gravity Simplicity Constraint as Surface Defect in Complex Chern-Simons Theory}

\author[1,2]{Muxin Han}

\author[1]{,\ \ Zichang Huang}

\affiliation[1]{Department of Physics, Florida Atlantic University, 777 Glades Road, Boca Raton, FL 33431-0991, USA}

\affiliation[2]{Institut f\"ur Quantengravitation, Universit\"at Erlangen-N\"urnberg, Staudtstr. 7/B2, 91058 Erlangen, Germany}

\emailAdd{hanm(AT)fau.edu, zhuang2014(AT)fau.edu } %

\abstract{The simplicity constraint is studied in the context of 4d spinfoam models with cosmological constant. We find that the quantum simplicity constraint is realized as the 2d surface defect in $\Slc$ Chern-Simons theory in the construction of spinfoam amplitudes. By this realization of simplicity constraint in Chern-Simons theory, we are able to construct the new spinfoam amplitude with cosmological constant for arbitrary simplicial complex (with many 4-simplices). The semiclassical asymptotics of the amplitude is shown to reproduce correctly the 4-dimensional Einstein-Regge action with cosmological constant term.
}


\arxivnumber{}

\begin{document}

\maketitle%\vspace{-7mm}

\section{Introduction}

There has been significant development recently on including cosmological constant in Loop Quantum Gravity (LQG) \cite{HHKR,HHKRshort,3dblockHHKR,hanSUSY,Han:2016dnt,curvedMink}\footnote{See e.g.\cite{book,review,review1,rovelli2014covariant,Perez2012} for reviews of Loop Quantum Gravity, including both the canonical and covariant formalisms.}. A new covariant formulation of LQG has been developed, and presented a nice relation between the covariant LQG in 4 dimensions and Chern-Simons theory on 3-manifold. In this new formalism, the spinfoam vertex amplitude is constructed by using the $\Slc$ Chern-Simons theory on 3-sphere with a Wilson graph. This new formulation using Chern-Simons theory evolves from the earlier formulation using quantum groups \cite{QSF,QSF1,NP}.

This work focuses on the spinfoam amplitude constructed from the new formalism. In particular, this work studies the quantum implementation of \emph{simplicity constraint} to the spinfoam amplitude in the presence of cosmological constant. It turns out that the simplicity constraint is realized as the 2d surface defect in $\Slc$ Chern-Simons theory used in constructing spinfoam amplitudes. By this realization of simplicity constraint in Chern-Simons theory, we are able to construct nonperturbatively the new spinfoam amplitude with cosmological constant for arbitrary simplicial complex (with many 4-simplices). The semiclassical asymptotics of the amplitude is shown to reproduce correctly the 4-dimensional Einstein-Regge action with cosmological constant term. 

In the classical Plebanski formulation, gravity in 4 dimensions is formulated by the topological BF theory and implementing the simplicity constraint. The simplicity constraint restricts the bivector $B$-field to be simple and relate to tetrad by $B^{IJ}=*(e^I\wedge e^J)$, which reduces BF action to the Palatini action of gravity. 

In the spinfoam formulation of covariant LQG, the simplicity constraint is quantized and imposed to partition function of quantum BF theory. In Engle-Pereira-Rovelli-Livine/Freidel-Krasnov (EPRL/FK) spinfoam model \cite{EPRL,FK}, a linear version of simplicity constraint is imposed in the spinfoam amplitude. Given a simplicial complex, the linear simplicity constraint states that for each tetrahedron $t$, the bivectors $B$-field smeared on its faces $B^{IJ}_f$ share the same time-like normal vector $N^I$. It is convenient to fix the time gauge that locally in each tetrahedron, the reference frame is chosen such that $N_I=(1,0,0,0)$. The time gauge breaks the local Lorentz symmetry to 3d rotation symmetry. The simplicity constraint then implies that all bivectors $B_f^{IJ}$ are spatial for each tetrahedron, and relate to the spatial normals of tetrahedron faces.

EPRL/FK spinfoam model is obtained by quantizing the above linear simplicity constraint and imposing \emph{weakly} to BF partition function \cite{generalize,DFLS,Dupuis2011,Speziale:2012nu}. The reason of imposing constraint weakly is that at the quantum level the components of linear simplicity constraint are not commutative. Strongly imposing the constraint results in that the solution space doesn't have enough degrees of freedom. Similar phenomena also happens in the Gupta-Bleuler formalism of quantizing electromagnetic field, and the covariant quantization of strings.  

The quantum simplicity constraint of EPRL/FK model guarantees that (1) the boundary degrees of freedom of spinfoam amplitude match precisely with the quantum 3d geometry emerging from canonical LQG. Namely, the boundary data of EPRL/FK amplitude are SU(2) spin-network data. (2) The semiclassical large spin asymptotics of the spinfoam amplitude reproduces correctly the Einstein-Regge action (without cosmological constant term) evaluated at simplicial geometries with flat 4-simplices \cite{CFsemiclassical,semiclassical,HZ,LowE,Han:2016fgh}. 

This work carries out the analysis of simplicity constraint for the spinfoam model with cosmological constant. The 4-dimensional spinfoam amplitude with cosmological constant is constructed by using $\Slc$ Chern-Simons theory on 3-manifold \cite{HHKR,HHKRshort,3dblockHHKR}\footnote{Following \cite{HHKR}, $\Slc$ Chern-Simons theory can be viewed to be equivalent to 4d BF theory with a cosmological constant term, when the 3d space where Chern-Simons lives is the boundary of the 4d space where BF theory lives. Schematically, the BF action with cosmological constant reads $S_{BF}=\int_{\sm_4}B^{IJ}\wedge *F_{IJ}+\frac{\L}{6}B^{IJ}\wedge *B_{IJ}$. Integrating out B-field leads to $S\sim \frac{1}{\L}\int_{\sm_4}F^{IJ}\wedge *F_{IJ}\sim \frac{i}{\L}\int_{\partial\sm_4}\tr(A\wedge\rmd A+\frac{2}{3}A\wedge A\wedge A)+c.c$, where $I,J=0,\cdots,3$ are Lorentz vector indices. $A=A^i\sig_i$ is the $\slc$-valued complex Chern-Simons connection.  See \cite{HHKR} for details in the presence of Barbero-Immirzi parameter.}. In this formalism, the local Lorentz symmetry of 4d spinfoam model is translated to the $\Slc$ gauge symmetry of Chern-Simons theory. The bivector $B_f^{IJ}$ are naturally exponentiated and given by the holonomy of flat connection traveling transversely around the Wilson line. The tetrahedron in 4d spinfoam model corresponds to the neighborhood of the vertex where 4 Wilson lines join (see FIG.\ref{gamma5} for the Wilson graph used for constructing 4-simplex amplitude).  

It is explained in Section \ref{SCCT} that the simplicity constraint and time gauge correspond to requiring that on the 4-holed sphere enclosing a 4-valent vertex, the gauge group of Chern-Simons is broken from $\Slc$ to SU(2). In the classical limit, the flat connections on 4-holed sphere are restricted to be SU(2). It is known that SU(2) flat connections on 4-holed sphere is in 1-to-1 correspondence to tetrahedron geometries with constant curvature \cite{curvedMink}. Thus imposing the simplicity constraint ensures the geometricity of tetrahedron at the classical level, similar to the situation of EPRL/FK model (with flat tetrahedron geometries). 

In Section \ref{QFSC}, we perform a quantization of the simplicity constraint, and define the constraint operators on the Hilbert space of SL(2,$\C$) Chern-Simons wave functions. Similar to the situation in EPRL/FK model, we find the constraint operators are noncommutative, which motivates us to rather impose a weaker version of constraints. We propose to use the master constraint technique \cite{Thiemann2006,Thiemann2006a,master}. The master constraint effectively reduces the Hilbert space to a subspace, whose wave functions are equivalent to SU(2) Chern-Simons wave functions. We might view the master constraint is a Hamiltonian, for which the SU(2) Chern-Simons wave functions on the 4-holed sphere are ground states, other $\Slc$ Chern-Simons states are created as excitations similar to harmonic oscillator. 

In addition, we find that the weak simplicity constraint is not unique. Indeed, the solution of the master constraint is a coherent state peaked at the phase space point which solves the classical simplicity constraint. We know that the coherent state which saturates the Heisenberg uncertainty is not unique, e.g. the squeezed coherent states. It turns out that different ways to define coherent states peaked at classical solutions of simplicity constraint correspond to different ways of weakly imposing simplicity constraint at the quantum level. %The vertex amplitude defined in \cite{HHKR} corresponds to a certain choices of coherent state.

In Section \ref{S3Gamma5}, we consider the graph complement 3-manifold $S^3\setminus\G_5$ similar to \cite{3dblockHHKR,HHKRshort}. We impose the quantum simplicity constraint and project the $\Slc$ Chern-Simons wave function to the space of solutions. The resulting wave function $\sz$ is proposed as a spinfoam 4-simplex amplitude with cosmological constant. We show that thanks to the simplicity constraint, the amplitude satisfies both (1) the boundary degrees of freedom match precisely with discrete 3d geometry data on the boundary of 4-simplex. The 3d geometry data is an analog of spin-network data (or semiclassically twisted geometry data) \cite{Han:2016dnt}. (2) The semiclassical asymptotics of the amplitude reproduce correctly the Einstein-Regge action with cosmological constant term on a constant curvature 4-simplex. The situation is an generalization of EPRL/FK model to include the cosmological constant.

In Section \ref{m3defect}, we generalize the analysis to arbitrary simplicial complex with many 4-simplices. The spinfoam amplitude on 4d simplicial complex is an $\Slc$ Chern-Simons theory on 3-manifold $\sm_3$ made by gluing copies of $S^3\setminus\G_5$. We find interestingly, the implementation of simplicity constraint corresponds to inserting 2d surface defects to $\Slc$ Chern-Simons theory on 3-manifold. The surface defects are inserted at the gluing interface (4-holed spheres) between pairs of $S^3\setminus\G_5$, i.e. they divide the entire 3-manifolds into copies of $S^3\setminus\G_5$. Each surface defect restricts the Chern-Simons states, which travel from one $S^3\setminus\G_5$ to another, to be solutions of simplicity constraint, i.e. to be equivalent to SU(2) Chern-Simons states. Because of understanding the simplicity constraint at the quantum level, we are able to formulate the spinfoam amplitude nonperturbatively on arbitrary simplicial complex, which improves the result in \cite{hanSUSY}. 
 
Because surface defects impose the quantum simplicity constraint, the two key properties of 4-simplex amplitude are generalized to the general spinfoam amplitude on simplicial complex. The boundary data are always 3d geometry data, so the amplitude describes the quantum history of 3d geometries. The semiclassical asymptotics reproduce correctly the Einstein-Regge action with cosmological constant term on the simplicial complex.

4d simplicial geometries emerge from critical points of spinfoam amplitude, where locally each 4-simplex is of constant curvature. Interestingly, the 3-manifold $\sm_3$ carrying Chern-Simons theory has a number of nontrivial cycles, each of which associates a torus cusp defect. The longitude holonomy along the B-cycle of torus cusp is noncontractible, since it associates to a noncontractible cycle of 3-manifold. It turns out that each noncontractible cycle corresponds to a triangle in 4d simplicial complex, and the noncontractible B-cycle holonomy corresponds to the nontrivial deficit angle hinged by the triangle. The 4d curvature is effectively created by the nontrivial cycles of 3-manifold $\sm_3$. 

In Section \ref{field}, we consider the field-theoretic description of the surface defect. We can define an operator insertion in the Chern-Simons path integral in terms of the continuous field theory variable. The 2d ``surface operator'' inserted in the path integral effectively implements the quantum simplicity constraint. In general, the defect of topological quantum field theory has certain dependence on the background metric, since it breaks the topological invariance to certain extend. A typical example is the framing dependence of Wilson line operators. Here the surface defect implementing the simplicity constraint also depends on a choice of metric on the 2-surface. Different choices of metrics in the field-theoretic context may be viewed as analogs of choosing different squeezed coherent states mentioned above. Thus different surface metrics for the surface defect correspond to different ways to implement weakly the quantum simplicity constraint. 

The semiclassical behavior is checked for the spinfoam amplitude in this field-theoretic formulation. The asymptotics again reproduce the Einstein-Regge action with cosmological constant on the entire simplicial complex. 

Although line defects have been widely studied in Chern-Simons theory, the results about surface defects (or domain-walls) are insufficient in the literature (some results are e.g. \cite{Kapustin:2010if,Armoni:2015jsa,Antillon:1997xr}). The surface defect appearing here has not been studied before. In Section \ref{SDF}, we investigate the surface defect by studying the propagating physical degrees of freedom on the defect 2-surface. As it is mentioned above, the surface defect reduces $\Slc$ Chern-Simons states to SU(2) in order to implement the simplicity constraint. On the defect where the gauge symmetry is broken, the previous gauge degrees of freedom become the physically propagating degrees of freedom. In other words, some additional propagating degrees of freedom have to be implemented in order to recover the original gauge symmetry on the defect. The standard example is the boundary of Chern-Simons theory, on which Wess-Zumino-Witten (WZW) model describes the propagating degree of freedom. We analyze the additional propagating degrees of freedom on the surface defect, which re-install the $\Slc$ gauge invariance to the model. We show that at least at the linearized level, the propagating field behaves as a 2d sigma model gauged by the Chern-Simons connection.

\section{Simplicity Constraint and Curved Tetrahedron}\label{SCCT}

In the spinfoam formulation without cosmological constant, the classical linear simplicity constraint requires that, given a flat tetrahedron $t$, each of the 4 face bivectors $B_f^{IJ}$ should be orthogonal to the time-like normal $N^I$ of the tetrahedron\footnote{$I,J=0,\cdots,3$ are vector indices of Lorentz group.}
\be
B^{IJ}_fN_I=0,\quad \forall f\subset \partial t.\label{BN}
\ee 
The time gauge may be chosen such that $N_I=(1,0,0,0)$, understood as a frame choice inside the tetrahedron. The frame can be located at any point inside the tetrahedron since the tetrahedron is flat. 

The choice of time gauge breaks the local Lorentz symmetry down to spatial rotation symmetry. We have for each bivector $B_f^{IJ}=B_f^{ij}$ where $i,j$ are 3d vector indices, and
\be
\Ar_f \hat{n}_f^i=\frac{1}{2}\eps^{ijk}(B_f)_{jk},
\ee 
where $\hat{n}$ is a unit space-like vector. Moreover because of the closure constraint 
\be
0=\sum_{f=1}^4 B_f^{IJ}=\sum_{f=1}^4 \Ar_f\hat{n}_f,
\ee
we know that the data $B^{IJ}_f$ satisfying simplicity constraint endow the geometry to the tetrahedron $t$, in which $\Ar_f$ is the face area and $\hat{n}_f$ is the unit face normal vector. 

\begin{figure}[htbp]
\centering\includegraphics[width=2in]{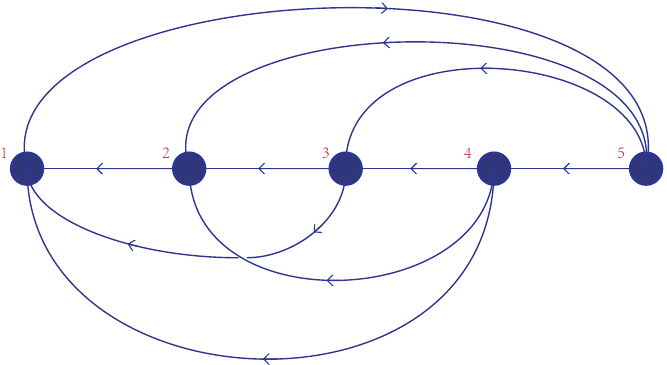}
\caption{$\Gamma_5$-graph embedded in $S^3$.}\label{gamma5}
\end{figure}

In the recent spinfoam models with cosmological constant, the 4d spinfoam 4-simplex amplitude is formulated as an SL(2,$\C$) Chern-Simons theory on $S^2$ with $\G_5$ Wilson graph defect (FIG.\ref{gamma5}) \cite{HHKR,3dblockHHKR,HHKRshort}. In this formulation, each tetrahedron of the 4-simplex relates to a 4-holed sphere $\cs$ enclosing a vertex of $\G_5$ graph. By the Chern-Simons equation of motion (in the semiclassical limit), the SL(2,$\C$) flat Chern-Simons connection on each 4-holed sphere gives a holonomy-version of closure constraint
\be
H_{4}H_3H_2H_1=1.\label{Hs}
\ee
Fixing a base point on $\cs$, $H_f$ is the holonomy of flat connection circling the $f$-th hole (each hole is dual to a tetrahedron face). The above formula can be viewed as a closure constraint generalizing $\sum_f B_f^{IJ}=0$ because each SL(2,$\C$) holonomy can be written as an exponential $H_f=\exp( \frac{\L}{3} B^{IJ}_f \cj_{IJ})$ where $\cj_{IJ}$ are Lorentz generators. When the cosmological constant $\L\to 0$, Eq.\Ref{Hs} implies the usual closure $\sum_f B_f^{IJ}=0$ by linearization.

When we apply the simplicity constraint Eq.\Ref{BN} and time gauge in this context, $B_f^{IJ}$ is again restricted to be spatial, thus
\be
H_f=\exp\lt( \frac{\L}{3} B^{IJ}_f \cj_{IJ}\rt)=\exp\lt(\frac{\L}{3}\Ar_f\hat{n}_f\cdot \vec{\t}\rt)\in\Su,\quad \vec{\t}=\frac{i}{2}\vec{\sig}\label{Hfan}
\ee
where $\vec{\sig}$ are Pauli matrices. Therefore the simplicity constraint and time gauge effectively reduce the structure group of Chern-Simons from $\Slc$ to $\Su$ on each 4-holed sphere. $\Slc$ flat connections are reduced to $\Su$. Eq.\Ref{Hs} becomes a product of SU(2) matrices.

It has been shown in \cite{HHKR,curvedMink} that the SU(2) flat connections on a 4-holed sphere $\cs$ are in 1-to-1 correspondence to the geometries of constant curvature tetrahedron, in which $\Ar_f$ in Eq.\Ref{Hfan} is the face area, $\hat{n}_f$ is the unit face normal. However since the tetrahedron is curved, a base point of tetrahedron has to be chosen in order to make sense of the frame choice for time gauge. Then $\hat{n}_f$ is the unit face normal vector located at the tetrahedron base point. 

The closure constraint Eq.\Ref{Hs} and the relation Eq.\Ref{Hfan} suggest that in the present of cosmological constant, the flux variable used in LQG are naturally exponentiated. The exponentiated flux variable has been recently studied in e.g.\cite{Han:2016dnt,Dittrich:2014wda,Sahlmann:2011rv}

The moduli space of flat SU(2) connection is of real dimension-6, which parametrizes all degrees of freedom for constant curvature tetrahedron geometries. The eigenvalues of SU(2) holonomies $H_f$ around the 4 holes relates to the 4 triangle areas of tetrahedron. It is shown in \cite{curvedMink} that the shapes of tetrahedron with fixed areas are parametrized by the flat connection coordinates $x,y\in \mathrm{U}(1)$. $x\in \mathrm{U}(1)$ relates to the diagonal length of a spherical 4-sided polygon, while $y\in \mathrm{U}(1)$ relates to the ``bending angle''.

In the moduli space of SL(2,$\C$) flat connections on $\cs$, the coordinates $x,y$ are known as Fenchel-Nielsen (FN) coordinates \cite{curvedMink,DGV,Han:2016dnt}\footnote{The FN coordinates is defined by cutting the 4-holed sphere $\cs$ into two 3-holed spheres. The flat connection on $\cs$ gives an $\Slc$ holonomy $h_x$ along the cut, whose eigenvalue is the FN complex length variable $x$. The FN twist variable $y$ is more technical to define. In a non-technical language, It comes from a holonomy $h_y$ of flat connection traveling from one 3-holed sphere to the other, which intersects transversely $h_x$. The diagonalization of $h_y$ gives the twist variable $y$. We refer to e.g. \cite{DGV,Han:2016dnt,3dblockHHKR} for a mathematically precise definition. }, while now $x,y\in \C\setminus\{0\}$ since they parametrize $\Slc$ flat connections. The symplectic structure of the moduli space induces that $x,y$ are canonically conjugate\footnote{The square on $y$ is conventional.}:
\be
\O=\frac{\rmd y^2}{y^2}\wedge\frac{\rmd x}{x}.
\ee 
Recall that the simplicity constraint reduces the flat connection on $\cs$ from $\Slc$ to $\Su$. In terms of the coordinates, the simplicity constraint implies 
\be
\mathrm{Re}(\ln x)=0,\quad \mathrm{Re}(\ln y)=0.\label{Re=0}
\ee
Namely, under the constraint, $x,y$ become U(1) numbers parametrizing the shape of tetrahedron. 

For the completeness, the simplicity constraint also restricts the eigenvalues of $H_f$ to be U(1) numbers as well, since they relate to face areas. But it turns out that these restrictions can be easily imposed at the quantum level. The only nontrivial task is to quantize the constraint Eq.\Ref{Re=0}, which we focus on in the following. For convenience, we often denote $X=\ln x$ and $P=\ln y^2$ in the following discussion.

\section{Quantization of Flat Connection and Simplicity Constraint}\label{QFSC}

We denote by $\cp_\cs$ the phase space of $\Slc$ flat connections on $\cs$ with fixed holonomy eigenvalue around each hole. $\cp_\cs$ is of 2 complex dimension. The coordinate on $\cp_\cs$ can be chosen to be $(x,y^2)$. The symplectic structure of $\Slc$ Chern-Simons theory reads
\be
\o_{k,s}&=&\frac{1}{4\pi}\lt(t\O+\bar{t}\bar{\O}\rt)\quad\quad\quad\quad\quad\quad t=k+is,\ \bar{t}=k-is\nonumber\\
&=&\frac{k}{2\pi}\lt(\rmd\,\mathrm{Re}P\wedge\rmd\,\mathrm{Re}X-\rmd\,\mathrm{Im}P\wedge\rmd\,\mathrm{Im}X\rt)-\frac{s}{2\pi}\lt(\rmd\,\mathrm{Re}P\wedge\rmd\,\mathrm{Im}X+\rmd\,\mathrm{Im}P\wedge\rmd\,\mathrm{Re}X\rt).
\ee

The quantization of phase space $\cp_\cs$ can be carried out in a similar way as in \cite{levelk}. $x=\exp X$ and $y^2=\exp P$ imply that $\mathrm{Im}X,\mathrm{Im}P$ are periodic with period $2\pi$. Weil's criterion of pre-quantization then requires $k\in\Z$. $s\in\R$ leads to $\o_{k,s}$ being real, so that the Chern-Simons theory is unitary with respect to a standard Hermitian struction \footnote{There is another unitary branch $s\in i\R$ via a nonstandard Hermitian structure \cite{Witten1991}}. 

As a convenient way to parametrize the complex Chern-Simons level, we write
\be
i s=k\frac{1-b^2}{1+b^2}\in i\R
\ee
with $|b|=1$. We can parametrize $x,\ y^2$ and their complex conjugates by
\be
x=\exp \frac{2\pi i}{k}\lt(-i b\mu-m\rt),&& \bar{x}=\exp \frac{2\pi i}{k}\lt(-i b^{-1}\mu+m\rt)\nonumber\\
y^2=\exp \frac{2\pi i}{k}\lt(-i b\nu-n\rt),&& \bar{y}^2=\exp \frac{2\pi i}{k}\lt(-i b^{-1}\nu+n\rt)
\ee
where $m,n\in\R$ are periodic $m\sim m+k,\ n\sim n+k$, $\mu,\nu$ are also real parameters. The Chern-Simons symplectic form $\o_{k,s}$ can be re-written in terms of new variables
\be
\o_{k,s}=\frac{2\pi}{k}\lt(\rmd \nu\wedge \rmd \mu-\rmd n\wedge \rmd m\rt)
\ee
The quantization of $\cp_\cs$ promotes the parameters $\mu,\nu,m,n$ to operators $\pmb{\mu},\pmb{\nu},\mathbf{m},\mathbf{n}$, whose non-vanishing commutation relation is
\be
\lt[\bfmu,\bfnu\rt]=\frac{k}{2\pi i},\quad \lt[\bfm,\bfn\rt]=-\frac{k}{2\pi i}.\label{ccr}
\ee
Or in terms of $\mathbf{x},\mathbf{y}^2$
\be
\mathbf{x}\mathbf{y}^2=q \mathbf{y}^2 \mathbf{x},\quad \bar{\mathbf{x}}\bar{\mathbf{y}}^2=\tilde{q} \bar{\mathbf{y}}^2 \bar{\mathbf{x}},\quad
q=\exp\frac{4\pi i}{t},\quad\tilde{q}=\exp\frac{4\pi i}{\bar{t}}.
\ee

The operator algebra is represented on the space of wave functions $f(\mu,m)$ of two variables. Here $\mu\in\R$ is continuous but $m\in\Z/k\Z$ is discrete. $m$ only takes integer value because both of the canonical conjugate variable $\bfm$ and $\bfn$ are periodic. Operators $\pmb{\mu},\pmb{\nu},\mathbf{m},\mathbf{n}$ are represented by
\be
&\pmb{\mu}f(\mu,m)=\mu f(\mu,m),&\quad  \pmb{\nu}f(\mu,m)=-\frac{k}{2\pi i}\partial_\mu f(\mu,m)\nonumber\\
&e^{\frac{2\pi i}{k}\bfm} f(\mu,m)=e^{\frac{2\pi i}{k}m} f(\mu,m),&\quad e^{\frac{2\pi i}{k}\bfn} f(\mu,m)=f(\mu,m+1).
\ee  

The simplicity constraint Eq.\Ref{Re=0} leads to the condition $\mu=\nu=0$ in the new parametrization. To quantize the constraint, one might naively impose the operator equations $\bfmu \psi=\bfnu \psi=0$ to the wave functions. However the naive operator equations trivialize the wave function since $\lt[\bfmu,\bfnu\rt]=\frac{k}{2\pi i}$. Therefore to realize the simplicity constraint at the quantum level, we have to impose a weaker version of the constraint. This fact makes it nontrivial for the quantum implementation of simplicity constraint. Here we choose to impose the operator equation
\be
(\bfmu-i\bfnu)\psi=0 \quad\Rightarrow\quad \psi_{\text{sol}}(\mu,m)=\exp\lt(-\frac{\pi\mu^2}{k}\rt)f(m).\label{qsimplicity}
\ee
where $f(m)$ is an arbitrary function on $\Z/k\Z$. Here the solution space is simply a $k$-dimensional vector space $\C^k$, being the Hilbert space of SU(2) Chern-Simons theory of level $k$. The simplicity constraint at quantum level reduces $\Slc$ Chern-Simons wave function to SU(2). 

As an equivalent way to impose the constraint, one may also consider to impose the ``master constraint'' $(\bfmu^2+\bfnu^2)\psi =0$ up to ``zero-point'' energy \footnote{See \cite{Thiemann2006,Thiemann2006a,master} for the idea of master constraint in canonical LQG. See \cite{EPRL,Speziale:2012nu} for the use of master constraint in spinfoam model to solve the simplicity constraint.}. The solution (the dependence on $\mu$) is simply the ground state of harmonic oscillator, the same as above. In this sense the states Eq.\Ref{qsimplicity} may be viewed as the ground states, while the full spectrum of $\Slc$ Chern-Simons states are created by the action of ``creation operator'' $(\bfmu+i\bfnu)$.

As we have seen, the constraint $\mu=\nu=0$ at the quantum level can only be satisfied weakly. The solution of the quantum constraint is a coherent state with peakedness at $\mu=\nu=0$. So $\mu=\nu=0$ is satisfied only in the semiclassical limit. It is known that the coherent state peaks at a phase space point is not unique. We may choose other squeezed coherent states, which still minimize the Heisenberg uncertainty. We introduce a squeezing parameter $w\in\R$, and impose $(\bfmu-i w^2\bfnu)\psi=0$ instead of Eq.\Ref{qsimplicity}, whose solution is 
\be
\psi^{(w)}_{\text{sol}}(\mu,m)=\exp\lt(-\frac{\pi\mu^2}{w^2k}\rt)f(m).
\ee
We may introduce a ``metric'' and define a ``squeezed'' master constraint $(w^{-2}\bfmu^2+ w^2\bfnu^2)$. The above squeezed coherent state satisfies $(w^{-2}\bfmu^2+ w^2\bfnu^2)\psi^{(w)}_{\text{sol}}=0$ up to the same zero-point energy as above (the zero-point energy is independent of $w$).

The squeezing parameter introduces an ambiguity to the solution of simplicity constraint at each 4-holed sphere $\cs$. The essential reason of the ambiguity is the noncommutativity of the simplicity constraints ${\bfmu}=0,{\bfnu}=0$. In the following Sections \ref{S3Gamma5} and \ref{m3defect}, we keep $w\neq 0$ as a free parameter, and focus on the construction of Chern-Simons theory with defects, as well as the geometrical reconstruction on $\sm_4$. We come back to the issue of ambiguity in Section \ref{field}.

\section{$\Slc$ Chern-Simons Theory on $S^3\setminus\G_5$}\label{S3Gamma5}

The partition function of $\Slc$ Chern-Simons theory on $S^3\setminus\G_5$ can be viewed as a wave function $Z_{S^3\setminus\G_5}(\l_\ell,\bar{\l}_\ell,x_\cs,\bar{x}_\cs)$ \cite{HHKRshort,3dblockHHKR}. The phase space of flat connections $\cp_{S^3\setminus\G_5}$ associated to the boundary of $S^3\setminus\G_5$ is of complex dimension 30. The boundary $\partial(S^3\setminus\G_5)$ is a closed 2-surface made of five 4-holed spheres $\cs$ connected by ten annuli $\ell$. A convenient system of coordinates is chosen to be the complex Fenchel-Nielsen (FN) coordinates $\l_\ell,\t_\ell$ for each annulus, and the $x_\cs,y_\cs^2$ (or $\mu_\cs,\nu_\cs,m_\cs,n_\cs$) coordinates for each 4-holed sphere $\cs$. Here $\l_\ell$ is the complex FN length, being the eigenvalue of meridian holonomy around the annulus $\ell$.

The implementation of simplicity constraint project the Chern-Simons wave function to the above solution space. The projection is done by the inner product: $|\,Z_{S^3\setminus\G_5}\,\rangle\to \lt|\psi_{\text{sol}}\rag\langle\psi_{\text{sol}}|\,Z_{S^3\setminus\G_5}\,\rangle$. The resulting wave function reads
\be
\mathscr{Z}_{S^3\setminus\G_5}\lt(\l_{\ell},\bar{\l}_{\ell},m_\cs\rt)=\int_{\R^5}\prod_\cs \rmd \mu_\cs\ Z_{S^3\setminus\G_5}\lt(\l_\ell,\bar{\l}_{\ell},\mu_\cs,m_\cs\rt)\prod_\cs\exp\lt(-\frac{\pi\mu^2_\cs}{w^2k}\rt).\label{Zint}
\ee
Here $\l_\ell$ is nothing but the eigenvalue of $H_f$ in Eq.\Ref{Hs}. The simplicity constraint about the eigenvalue of $H_f$ can be easily implemented by restricting $\l_\ell\in U(1)$ in the wave function. 

Now the wave function $\sz$ only depends on the data of SU(2) flat connections on Riemann surface $\Sig_6=\partial S^3\setminus\G_5$. It has been shown that the SU(2) flat connections on Riemann surface parametrizes the twisted geometry on 3d discrete space \cite{Han:2016dnt}. Thus $\sz$ is indeed qualified to be a quantum amplitude describing the evolution of 3d geometry. For the relation with spin-network data, it is explaned in a moment that $\l_{\ell}$ relates to the spins $j_\ell$. $m_{\cs}\in \Z/k\Z$ quantizes SU(2) flat connections on 4-holed sphere, thus essentially is a label of the conformal blocks (or equivalently, 4-valent intertwiners of quantum group $\Su_q$ with $q$ root of unity) \cite{Elitzur:1989nr}.

We consider the semiclassical limit of the resulting $Z_{S^3\setminus\G_5}(\l_{\ell},m_\cs)$ as $k,s\to\infty$. Comparing the semiclassical limit to the commutators Eq.\Ref{ccr} motivates us to rescale $\mu_\cs,\nu_\cs,m_\cs,n_\cs$ by
\be
\mu_\cs\mapsto \frac{k}{2\pi}\mu_\cs,\quad \nu_\cs\mapsto\frac{k}{2\pi}\nu_\cs,\quad m_\cs\mapsto \frac{k}{2\pi}m_\cs,\quad n_\cs\mapsto \frac{k}{2\pi}n_\cs\label{rescale}
\ee
After rescaling, $m_\cs,n_\cs$ become continuous periodic variables as $k\to\infty$. 

The semiclassical behavior of $Z_{S^3\setminus\G_5}(\l_\ell,\bar{\l}_\ell,\mu_\cs,m_\cs)$ is known as \cite{DGLZ,analcs} ($\cdots$ stands for the quantum corrections)
\be
&&Z_{S^3\setminus\G_5}\lt(\l_\ell,\bar{\l}_\ell,\mu_\cs,m_\cs\rt)\nonumber\\
&=&\sum_\a \exp\lt\{ i\int^{(\l_\ell,\bar{\l}_\ell,\mu_\cs,m_\cs)}_{c\subset\cl_\a}\lt[ \frac{t}{4\pi} \sum_\ell\ln \t_\ell\frac{\rmd \l'_\ell}{\l'_\ell}+\frac{\bar{t}}{4\pi} \sum_\ell\ln \bar{\t}_\ell\frac{\rmd \bar{\l}'_\ell}{\bar{\l}'_\ell}+\frac{k}{2\pi}\sum_\cs\lt(\nu_\cs \rmd\mu'_\cs+n_\cs\rmd m'_\cs\rt)\rt]+\cdots\rt\}.
\ee
The moduli space of flat connections on $S^3\setminus\G_5$, $\cm_{flat}(S^3\setminus\G_5,\Slc)\simeq\cl$ is understood as the Lagrangian submanifold of the phase space. $\cl_\a$ is the branch of $\cl$ associated to the flat connection $\a$ on $S^3\setminus\G_5$. $\cl$ can be represented as a set of polynomial equations in symplectic coordinates, whose expressions have been derived in \cite{hanSUSY}. The quantity on the exponential is an integral of the Liouville 1-form associated to $\o_{k,s}$ along a contour $c$ in $\cl_\a$.

Insert the asymptotic express of $Z_{S^3\setminus\G_5}\lt(\l_\ell,\bar{\l}_\ell,\mu_\cs,m_\cs\rt)$ in the integral Eq.\Ref{Zint},
\be
\sz_{S^3\setminus\G_5}\lt(\l_{\ell},\bar{\l}_{\ell},m_\cs\rt)
=\sum_\a \int_{\R^5}\rmd \mu_\cs\, \exp\lt[ S_\a\lt(\l_\ell,\bar{\l}_\ell,\mu_\cs,m_\cs\rt) +\cdots\rt].\label{Zint1}
\ee
where $S_\a\lt(\l_\ell,\bar{\l},\mu_\cs,m_\cs\rt)$ reads
\be
S_\a= i\int^{(\l_\ell,\bar{\l}_\ell,\mu_\cs,m_\cs)}_{c\subset\cl_\a}\lt[ \frac{t}{4\pi} \sum_\ell\ln \t_\ell\frac{\rmd \l'_\ell}{\l'_\ell}+\frac{\bar{t}}{4\pi} \sum_\ell\ln \bar{\t}_\ell\frac{\rmd \bar{\l}'_\ell}{\bar{\l}'_\ell}+\frac{k}{2\pi}\sum_\cs\lt(\nu_\cs \rmd\mu'_\cs+n_\cs\rmd m'_\cs\rt)\rt]-\sum_{\cs}\frac{ k\mu^2_\cs}{4\pi w^2}.
\ee

In the semiclassical limit, the $\mu_\cs$-integral Eq.\Ref{Zint1} localizes asymptotically at the critical points, i.e. the solutions of critical equations $\partial S_\a/\partial{\mu_\cs}=\mathrm{Re}(S_\a)=0$. The critical equations are easy to derive: 
\be
iw^2\nu_\cs-\mu_\cs=\mu_\cs=0\quad\Rightarrow\quad \mu_\cs=\nu_\cs=0,\label{munu=0}
\ee
where we see that the critical equation $\partial S_\a/\partial{\mu_\cs}=0$ is a classical version of the quantum simplicity constraint Eq.\Ref{qsimplicity}. The critical equations imply the simplicity constraint, thus require the flat connections on all $\cs$ to be SU(2).

The condition $\mu_\cs=\nu_\cs=0$ simultaneously  may not be satisfied for generic branches $\cl_\a$ of the Lagrangian submanifold. However it has been shown in \cite{3dblockHHKR} that there exists exactly 2 branches $\cl_{\a_{4d}}$ and $\cl_{\tilde{\a}_{4d}}$, where $\nu_\cs(\mu_\cs=0)=0$ can be satisfied. The $\Slc$ flat connection $\a_{4d}$ on $S^3\setminus\G_5$ equivalently describes the geometry of a nondegenerate 4-simplex with constant curvature. The other flat connection $\tilde{\a}_{4d}$ is referred to as the ``parity partner'', which corresponds to the same 4-simplex geometry as $\a_{4d}$, but with opposite 4d orientation. 

Those $\a$ whose $\cl_\a$ doesn't consistent with $\mu_\cs=\nu_\cs=0$ only give exponentially suppressed contributions to $\sz_{S^3\setminus\G_5}\lt(\l_{\ell},\bar{\l}_{\ell},m_\cs\rt)$ in Eq.\Ref{Zint1}. Therefore
\be
\sz_{S^3\setminus\G_5}\lt(\l_{\ell},\bar{\l}_{\ell},m_\cs\rt)= e^{S_{\a_{4d}}\lt(\l_{\ell},\bar{\l}_{\ell}\rt)+\cdots}+ e^{S_{\bar{\a}_{4d}}\lt(\l_{\ell},\bar{\l}_{\ell}\rt)+\cdots}
\ee
where $S_{\a_{4d}}$ reads\footnote{We have choose the integration contour such that the flat connections on the contour all correspond to the 4d geometries. Therefore the Schl\"afli identity can be used in the derivation (see \cite{3dblockHHKR} for details). The contour is in the plane with $\mu_\cs=0$.}
\be
S_{{\a}_{4d}}=i\int^{(\l_\ell,\bar{\l}_\ell,m_\cs)}_{c\subset\cl_{{\a}_{4d}}}\lt[ \frac{t}{4\pi} \sum_\ell\ln \t_\ell\frac{\rmd \l'_\ell}{\l'_\ell}+\frac{\bar{t}}{4\pi} \sum_\ell\ln \bar{\t}_\ell\frac{\rmd \bar{\l}'_\ell}{\bar{\l}'_\ell}+\frac{k}{2\pi}\sum_\cs n_\cs\rmd m'_\cs\rt],\label{integral1}
\ee
The integral in $S_{\a_{4d}}$ has been reduced to be of the same form as the one treated in \cite{3dblockHHKR,HHKRshort}. 

To compute $S_{{\a}_{4d}}$, we use the geometrical interpretation of flat connections and the FN coordinates in terms of constant curvature 4-simplex geometries. This geometrical interpretation has been studied expensively in \cite{HHKR,3dblockHHKR}. The 10 annuli $\ell$ are of 1-to-1 correspondence to the 10 triangles of the 4-simplex. By the correspondence between 4-simplex geometry and flat connection on $S^3\setminus\G_5$, the complex FN length $\l_\ell$ relates to the area of the triangle $\mathbf{a}(\ff_{\ell})$. The dihedral angle $\Theta(\ff_\ell)$ hinged by the triangle $\ff_\ell$ corresponding to $\ell$ relates to the complex FN twist $\t_\ell$. Explicitly,
\be
\l_{\ell}=\exp\lt[-\frac{i\L}{6}\mathbf{a}(\ff_{\ell})+\pi i\fs_{\ell}\rt],\quad \t_\ell=\exp\lt[-\sgn(V_4)\,\Theta(\ff_\ell)\rt]\label{lamdaarea}
\ee 
where $\fs_\ell\in\{0,1\}$ parametrizes the lifts from $\PSlc$ to $\Slc$. $\sgn(V_4)$ is the 4d orientation of the 4-simplex, which takes different values at $\a_{4d}$ and $\bar{\a}_{4d}$.

Insert Eq.\Ref{lamdaarea} in the integral Eq.\Ref{integral1}, the integrand becomes proportional to $\sum_{\ell=1}^{10}\Theta(\ff_\ell)\rmd \mathbf{a}(\ff_{\ell})$ except the last term in Eq.\Ref{integral1}. Because all data $\Theta(\ff_\ell), \mathbf{a}(\ff_{\ell})$ associates to a geometrical 4-simplex, and satisfy the Schl\"afli identity \cite{eva,Haggard:2014gca}
\be
\sum_{\ell=1}^{10}\mathbf{a}(\ff_{\ell})\rmd \Theta(\ff_\ell)=\L \rmd |V_4 |
\ee
where $V_4$ is the volume of the constant curvature 4-simplex, $\sum_{\ell=1}^{10}\Theta(\ff_\ell)\rmd \mathbf{a}(\ff_{\ell})$ is a total derivative:
\be
\sum_{\ell=1}^{10}\Theta(\ff_\ell)\rmd \mathbf{a}(\ff_{\ell})=\rmd S_{\text{Regge},\L},\quad S_{\text{Regge},\L}=\sum_{\ell}\mathbf{a}(\ff_{\ell})\,\Theta(\ff_\ell)-\L |V_4|
\ee
$S_{\text{Regge},\L}$ is the Regge action on a single 4-simplex with cosmological constant term.

The last term in Eq.\Ref{integral1} contribute the same between $\a_{4d}$ and $\bar{\a}_{4d}$ \cite{3dblockHHKR}. To remove this overall term in the asymptotics, we may consider a coherent state peaked at the phase space point $\mathring{m}_\cs,\mathring{n}_\cs$, which behaves as the following when $k\to\infty$,
\be
\phi^{(k)}_{\mathring{m},\mathring{n}}\lt(m_\cs\rt)\sim e^{-\frac{k}{4\pi}\sum_\cs\lt(m_\cs -\mathring{m}_\cs\rt)^2-\frac{ik}{2\pi}\sum_\cs\mathring{n}_\cs m_\cs}.\label{CohState}
\ee
A candidate of $\phi^{(k)}$ can be chosen as a product of Jacobi Theta functions (see Eq.(4.19) of \cite{3dblock}) to respect the periodicity of $m_\cs$. As $k\to\infty$, the quantity
\be
\sum_{m_\cs}\sz_{S^3\setminus\G_5}\lt(\l_{\ell},\bar{\l}_{\ell},m_\cs\rt)\phi^{(k)}_{\mathring{m},\mathring{n}}\lt(m_\cs\rt)\label{Q}
\ee
gives the critical equation of $m_\cs$:
\be
m_\cs=\mathring{m}_\cs, \quad n_{\cs}=\mathring{n}_\cs.
\ee
At the critical point, the last term in Eq.\Ref{integral1} cancels the second term on the exponential in $\phi^{(k)}$.

As a result, Eq.\Ref{Q} behaves asymptotically as
\be
 e^{\frac{i}{\ell_P^2} S_{\text{Regge},\L}+\cdots}+ e^{-\frac{i}{\ell_P^2}S_{\text{Regge},\L}+\cdots}\label{Zregge}
\ee
where $\ell_P^2=\lt|\frac{12\pi}{s\L}\rt|$.

The above asymptotics reproduce the result in \cite{HHKR,3dblockHHKR,HHKRshort}. The previous asymptotic results have been obtained either by pick up semiclassically the branches $\a_{4d},\bar{\a}_{4d}$, or by a certain ansatz of Wilson graph operator. However here we obtain the result by a systematic study of the simplicity constraint at the quantum level, and project the partition function onto the space of quantum solutions. The method used here is especially useful when generalizing the amplitude to many 4-simplices.

\section{$\Slc$ Chern-Simons theory on $\sm_3$ with Surface Defect}\label{m3defect}

\begin{figure}[h]
\begin{center}
\includegraphics[width=10cm]{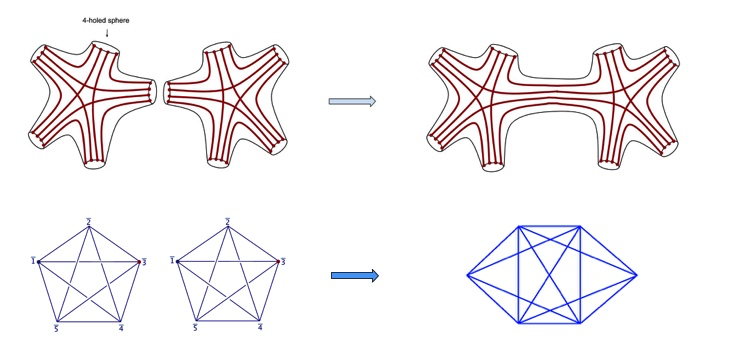}
\caption{$\sm_3$ is obtained by gluing a number of $S^3\setminus\G_5$, each of which corresponds to a 4-simplex in $\sm_4$. The gluing of $S^3\setminus\G_5$'s is deduced from the gluing of 4-simplices in $\sm_4$. In drawing the 3-manifold $S^3\setminus\G_5$ and $\sm_3$, we imagine to view $S^3\setminus\G_5$ from 4d and suppress 1 dimension. The 3-manifold $S^3\setminus\G_5$ has five geodesic boundary components as 4-holed spheres, coming from removing the neighborhood of five vertices of $\G_5$. It has ten cusp boundary components as ten annuli, coming from removing the neighborhood of ten edges of $\G_5$. The red curves are the annuli connecting 4-holed spheres. Two $S^3\setminus\G_5$ can be glued through a pair of 4-holed spheres, via a certain identification of holes. Each 4-holed sphere as the gluing interface corresponds to a tetrahedron shared by two 4-simplices in $\sm_4$. Each hole of the 4-holed sphere (or each tunnel traveling thought the 4-holed sphere) corresponds to a triangle in the shared tetrahedron. }
\label{m3andm4}
\end{center}
\end{figure}

The correspondence between $\Slc$ flat connection on 3-manifold and 4d geometry can be generalized to arbitrary 4d simplicial manifold $\sm_4$. The corresponding 3-manifold $\sm_3$ corresponding to $\sm_4$ can be constructed by gluing copies of $S^3\setminus\G_5$ (see FIG.\ref{m3andm4}). The number of glued $S^3\setminus\G_5$ coincides with the number of 4-simplices in $\sm_4$. The gluing interface between a pair of $S^3\setminus\G_5$ is always a 4-holed sphere $\cs$.

To construct the partition function $\sz_{\sm_3}$ on $\sm_3$, we simply product the resulting partition function $\sz_{S^3\setminus\G_5}\lt(\l_{\ell},\bar{\l}_{\ell},m_\cs\rt)$ (reduced by the simplicity constraint), then identify and sum over the data $m_\cs$ associated to the gluing interfaces $\cs$. So we obtain a state-sum model.
\be
\sz_{\sm_3}\lt(\l_{\ell},\bar{\l}_{\ell}\rt)=\sum_{m_\cs\in \Z/k\Z}\,\prod_{S^3\setminus\G_5}Z_{S^3\setminus\G_5}\lt(\l_{\ell},\bar{\l}_{\ell},m_\cs\rt).\label{Zsm}
\ee
In this formula, $m_\cs\in\Z/k\Z$ is the one before the rescaling Eq.\Ref{rescale} in the semiclassical analysis. In general the resulting $\sz_{\sm_3}$ may also depend on some leftover $m_\cs$'s, in case that $\sm_3$ after gluing still has geodesic boundary components $\cs$. 

The simplicity constraint has been implemented at the gluing interfaces $\cs$. The constraint project the quantum states on $\cs$ of $\Slc$ Chern-Simons theory to the space of solutions Eq.\Ref{qsimplicity}, which is essentially the state space of SU(2) Chern-Simons theory. Therefore the simplicity constraint introduces the defects to $\Slc$ Chern-Simons theory. The defects is localized at the interfaces $\cs$ where a pair of $S^3\setminus\G_5$ are glued. The defects are supported on 2-surfaces $\cs$ embedded in $\sm_3$. The effect of the defect is that $\Slc$ Chern-Simons theory reduces to SU(2) at the 2-surface.

Schematically, the surface defect may be understood via the insertions of certain ``surface operators'' in $\Slc$ Chern-Simons theory, i.e. we write $\sz_{\sm_3}\lt(\l_{\ell},\bar{\l}_{\ell}\rt)$ as a functional integral
\be
\sz_{\sm_3}\lt(\l_{\ell},\bar{\l}_{\ell}\rt)=\int DAD\bA\ e^{\frac{it}{8\pi}\int_{\sm_3}\tr(A\rmd A+\frac{2}{3}A^3)+\frac{i\bar{t}}{8\pi}\int_{\sm_3}\tr(\bA \rmd \bA+\frac{2}{3}\bA^3)}\prod_{\cs}\co_\cs\lt[A,\bar{A}\rt]\label{insertion}
\ee
The insertions $\co_\cs$, located at the gluing interfaces $\cs$, play the role of the projections $\lt|\psi_{\text{sol}}\rag\langle\psi_{\text{sol}}|$. The further discussion of this operator $\co_\cs\lt[A,\bar{A}\rt]$ is given in Section \ref{field}. 

We consider the semiclassical behavior of the state-sum $\sz_{\sm_3}$ as $k,s\to\infty$. We again perform the rescaling for $m_\cs$ by Eq.\Ref{rescale}. Then we see that as $k\to\infty$ the sum over $m_\cs$ in Eq.\Ref{Zsm} approximates an integral over $S^1$. The semiclassical asymptotics can be again studied by stationary phase approximation, similar to the analysis of $Z_{S^3\setminus\G_5}$. In addition to the critical equations Eq.\Ref{munu=0}, we have one more critical equation at each gluing interface $\cs$, because of the integration of $m_\cs$.
\be
n_{\cs}+n'_\cs=0
\ee
where $n_\cs$ comes from the flat connection on the $S^3\setminus\G_5$ on the left of $\cs$, and $n'_\cs$ comes from the $S^3\setminus\G_5$ on the right of $\cs$.

Semiclassically $n_{\cs}=-n'_\cs$ identifies the SU(2) flat connections on the interface $\cs$ from the left and right $S^3\setminus\G_5$'s ($m_\cs$ has been identified). The minus sign reflects the opposite orientations on $\cs$ in the gluing. Thus the $\Slc$ flat connections on the copies of $S^3\setminus\G_5$ are glued to become a flat connection on the entire $\sm_3$.  

Let's first consider $\sm_3$ is obtained by gluing 2 copies of $S^3\setminus\G_5$ through a pair of 4-holed spheres $\cs,\cs'$, the fundamental group $\pi_1(\sm_3)$ is given by two copies of $\pi_1(S^3\setminus\G_5)$ modulo the identification of generators on $\cs$ and $\cs'$ ($\pi_1(\cs)\simeq\pi_1(\cs')$ with the isomorphism denoted by $\ci$). $\pi_1(\sm_3)$ is isomorphic to the fundamental group of a 1-skeleton $\pi_1(\mathrm{sk}(\sm_4))$ from the 4d polyhedron $\sm_4$ obtained by gluing a pair of 4-simplices. However here the 1-skeleton $\mathrm{sk}(\sm_4)$ includes the edges of the tetrahedron shared by the pair of 4-simplices (FIG.\ref{m3andm4}).

Given two flat connections $A,A'$ as representations $\pi_1(S^3\setminus\G_5)\to \Slc$ modulo conjugation, they are glued and give a flat connection $\sa$ on $\sm_3$ when they induce the same representation to $\pi_1(\cs)$ and $\pi_1(\cs')$ (i.e. $A=A'\circ \ci$). We reduce the flat connection on $\cs,\cs'$ to be SU(2), and consider $A,A'$ corresponds to 2 constant curvature 4-simplices $\Fs,\Fs'$. When $A,A'$ glue to $\sa$ on $\sm_3$, they induce the same SU(2) representation (modulo conjugation) to $\pi_1(\cs)$ and $\pi_1(\cs')$. The SU(2) flat connection reconstructs a unique geometrical tetrahedron of constant curvature. The constant curvature tetrahedron belongs to both $\Fs,\Fs'$, and implies $\Fs,\Fs'$ are of the same constant curvature. Therefore the flat connection $\sa$ on $\sm_3$ effectively glues a pair of constant curvature 4-simplices $\Fs,\Fs'$, and determines a 4-dimensional simplicial geometry on $\sm_4$. The procedure can be continued to arbitrary $\sm_3=\cup_{i=1}^N (S^3\setminus\G_5)$. For each simplicial 4-manifold $\sm_4$, the corresponding $\sm_3$ can be constructed as in FIG.\ref{m3andm4}. A class of flat connections $\sa$ on $\sm_3$ can be obtained by gluing flat connections on $S^3\setminus\G_5$. Each $\sa$ determines a 4d simplicial geometry $(\sm_4,g)$ obtained by gluing $N$ 4-simplices with the same constant curvature. When the simplicial complex $\sm_4$ is sufficiently refined, arbitrary smooth geometries can be approximated by the simplicial geometries.   

The gluing of flat connections gives extra constraint on $A,A'$ as well as the boundary data $\l_\ell$. It is possible that a set of $\l_{\ell}$ doesn't lead to any flat connection on $\sm_{3}$ corresponding to 4d simplicial geometry. In that case, we say the areas relating to $\l_\ell$ are \emph{non-Regge-like}, otherwise we say the areas are \emph{Regge-like}.

\begin{figure}[h]
\centering\includegraphics[width=3.2in]{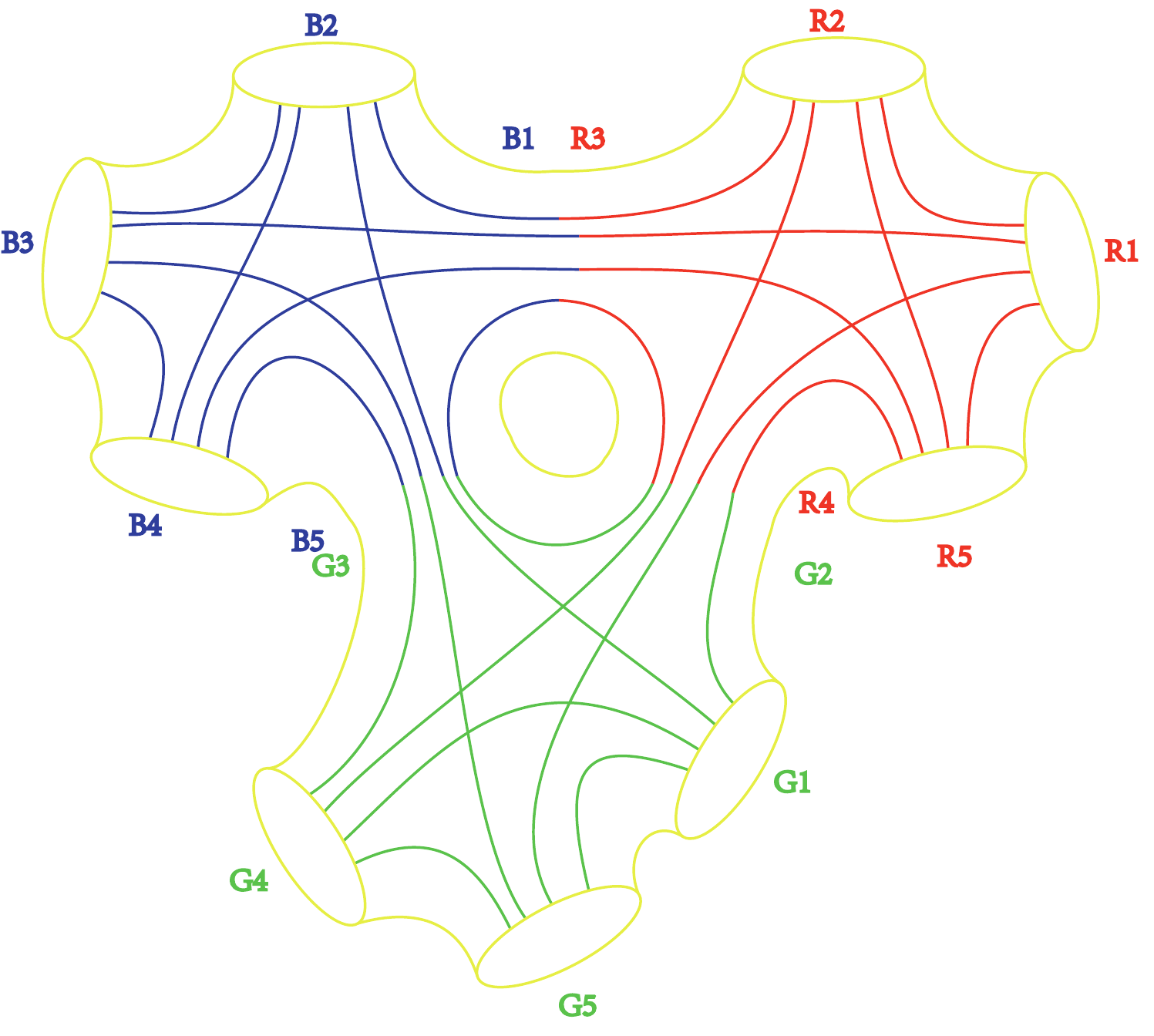}
\caption{This picture shows a result that we glue three $\mathcal{S}^3\backslash \Gamma_5$. The yellow shell outside indicate the ambient 3-manifold $\mathfrak{X}_3$. Four-holed spheres $B1$ and $R3$ are shared boundary between blue ${S}^3\backslash \Gamma_5$ and red ${S}^3\backslash \Gamma_5$. Similarly, $(B5, G3)$ and $(R4,G2)$ are blue-green shared boundary and red-green shared boundary respectively. At the center of the picture, there is a non-contractible cycle, which made $\pi_1(\mathfrak{X}_3)$ nontrivial. There is a closed tunnel with three different colors at the center corresponds to an internal triangle shared by three 4-simplices in $\sm_4$.}\label{fig:7}
\end{figure}

In general, $\sm_3$ can be viewed as the complement of (the open neighborhood of) certain graph $\G$ in an ambient closed 3-manifold $\Fx_3$. Generically $\mathfrak{X}_3$ is not $\mathcal{S}^3$. It is manifest as an example in Figure \ref{fig:7}, when we glue three $S^3\setminus\G_5$. $\mathfrak{X}_3$ has a non-contractible cycle which are generated by the gluing procedure. In another word, the fundamental group $\pi_1(\mathfrak{X}_3)$ is non-trivial. In the case of FIG.\ref{fig:7}, the non-contractible cycle of $\Fx_3$ associates a closed tunnel from connecting a number of annuli in $S^3\setminus\G_5$. In general each closed tunnel always goes along a non-contractible cycle in $\Fx_3$. The tunnel gives a torus boundary $T^2$ of $\sm_3$. Following the correspondence between $\sm_3$ and $\sm_4$, it is not hard to see that each torus boundary $T^2$ corresponds to an internal triangle shared by a number of 4-simplices.

The flat connection $\sa$ gives the commutative meridian (A-cycle) and longitude (B-cycle) holonomies on each $T^2$. The commutativity implies the two holonomies can be simultaneously diagonalized. The eigenvalue $\l_{T^2}$ of the meridian holonomy equals to the annulus meridian holonomy eigenvalue $\l_\ell$ for all $\l_\ell$ building $T^2$. From the correspondence between $\sa$ and simplicial 4-geometry, it is not hard to see that $\l_{T^2}$ relates to the area $\mathbf{a}(\ff_{T^2})$ of the internal triangle $\ff_{T^2}$
\be
\l_{T^2}^2=\exp\lt[-\frac{i\L}{3}\mathbf{a}(\ff_{T^2})\rt].\label{lllll}
\ee 
The eigenvalue $\t_{T^2}$ of the longitude holonomy can be obtained by a product of FN twists $\t_\ell$ for all $\l_\ell$ building $T^2$. It has been computed in \cite{hanSUSY}. When we connect a pair of annuli $\ell_1,\ell_2$ in $\Fx_3$, the FN twist of the connected annuli $\ell_1\cup\ell_2$ is a product of the FN twists of $\ell_1$ and $\ell_2$, and the same relation also holds for $y^2_\ell$:
\be
\t_{\ell_1\cup\ell_2}=\t_{\ell_1}\t_{\ell_2},\quad y^2_{\ell_1\cup\ell_2}=y^2_{\ell_1}y^2_{\ell_2}.
\ee
When a number of annuli are connected to form a $T^2$, $\t_{T^2}=y_{\ell_1\cup\cdots\cup\ell_n}$ is the eigenvalue of the longitude holonomy, which is a product of $y_{\ell_1},\cdots,y_{\ell_n}$. We also have $\t_{T^2}^2=\t_{\ell_1\cup\cdots\cup\ell_n}$. Because of the relation between $y_{\ell}$ and 4-simplex hyper-dihedral angle in Eq.\Ref{lamdaarea}, $\t_{T^2}$ relates to the deficit angle $\eps(\ff_{T^2})$ hinged by the internal triangle $\ff_{T^2}$
\be
\t_{T^2}=e^{-\half\sgn(V_4)\,\eps(\ff_{T^2})-\frac{i}{2}\pi\eta(\ff_{T^2})},\ \ \ \ \text{where}\ \ \ \ \eps(\ff_{T^2})=\sum_{\Fs,\ \ff_{T^2}\subset \Fs}\Theta_\Fs(\ff_{T^2}).\label{tauT2}
\ee 
when $\sgn(V_4)$ is a constant for all 4-simplices sharing $\ff_{T^2}$. It is shown in \cite{hanSUSY} that $\eta(\ff_{T^2})$ is an index taking values in $\{0,1\}$. $\eta(\ff_{T^2})=0$ everywhere on $\sm_4$ means that the 4d spacetime is globally time-oriented.

The gluing of 4-simplices via gluing $S^3\setminus\G_5$ does not put any constraint on the orientation $\sgn(V_4)$ of each 4-simplex. There are $2^N$ flat connections on $\sm_4$ corresponding to the same geometry on $\sm_4$ but of different local orientations. Among them there are a pair of flat connections give the globally oriented geometry on $\sm_4$.

We again label by $\a_{4d}$ the flat connection on $\sm_3$ which corresponds to 4d simplicial geometry, which is globally oriented ($\sgn(V_4)$ is constant) and globally time-oriented ($\eta(\ff_{T^2})$ vanishes constantly). The contribution from each $\a_{4d}$ to $Z_{\sm_3}$ asymptotically behaves as $Z_{\sm_3}\sim \exp S_{\a_{4d}}$, when $k,s\to\infty$, where
\be
S_{{\a}_{4d}}&=&i\int^{(\l,\bar{\l},m)}_{c\subset\cl_{{\a}_{4d}}}\Bigg[ \frac{t}{4\pi} \lt(\sum_{T^2\subset\partial\sm_3}\ln \t_{T^2}\frac{\rmd {\l'}_{T^2}^2}{{\l'}_{T^2}^2}+\sum_{\ell\subset\partial\sm_3}\ln \t_\ell\frac{\rmd \l'_\ell}{\l'_\ell}\rt)+\frac{\bar{t}}{4\pi}\lt(\sum_{T^2\subset\partial\sm_3}\ln \bar{\t}_{T^2}\frac{\rmd {\bar{\l'}^2_{T^2}}}{{\bar{\l'}^2_{T^2}}}+\sum_{\ell\subset\partial\sm_3}\ln \bar{\t}_\ell\frac{\rmd \bar{\l'}_\ell}{\bar{\l'}_\ell}\rt)\nonumber\\
&& +\frac{k}{2\pi}\sum_{\cs\subset\partial\sm_3} n_\cs\rmd m'_\cs\Bigg],\label{S4d0}
\ee
The type of integral has been computed in \cite{hanSUSY}. The method of computation is similar to Eq.\Ref{integral1}. Key steps are again using the geometrical interpretations Eqs.\Ref{tauT2} and \Ref{lllll}, as well as the Schl\"afli identity for each 4-simplex. The result gives the Einstein-Regge action on the simplicial complex $\sm_4$, up to some additional boundary terms which correspond to the overall phase of the wave function (see Section 6 in \cite{hanSUSY} for details):
\be
S_{{\a}_{4d}}=-\frac{i s\L\sgn(V_4)}{12\pi}\lt(\sum_{\ff}\mathbf{a}(\ff)\,\eps(\ff)-\L \sum_{\Fs}\lt|V_4(\Fs)\rt|\rt)+\frac{ik\L}{3}\sum_{\ff}N(\ff)\, \mathbf{a}(\ff)  +\frac{ik}{2\pi}\int^{(\l,\bar{\l},m)}_{\cl_{{\a}_{4d}}}\sum_{\cs\subset\partial\sm_3} n_\cs\rmd m'_\cs \label{S4d}
\ee
where we have neglected the integration constant. To make the formula short, $\eps(\ff)$ here denotes the deficit angle for internal $\ff$ or the dihedral angle for boundary $\ff$. We read the coefficient in front of Regge action to be the (inverse) Planck scale in 4d
\be
\ell_P^2=\lt|\frac{12\pi}{s\L}\rt|.
\ee
$N(\ff)$ indicates the leading order of $S_{{\a}_{4d}}$ is a multi-valued function since it comes from integrating the logarithmic function (see \cite{3dblock} for an interpretation). However if there is a quantization of area 
\be
\frac{\L}{3}\sum_{\ff}N(\ff)\, \mathbf{a}(\ff)\in 2\pi\Z
\ee
The asymptotics of $\exp S_{\a_{4d}}$ doesn't depend on the choice of branches $N(\ff)$. This area-quantization condition has been treated in \cite{3dblockHHKR}. It is fulfilled when the boundary condition $\l_\ell$ comes from the $\Slc$ Wilson lines in $\Fx_3$ labelled by unitary irreps $(2j_\ell,2\g j_\ell)$ where $j_\ell\in \mathbb{N}/2$ and $\g=s/k$ is a universal constant. The area relates the representation label by $\mathbf{a}(\ff)=\g j_\ell$ with the correspondence between $\ff$ and $\ell$.

The last term in Eq.\Ref{S4d} only relates to the boundary of $\sm_3$ or $\sm_4$. If we fix the boundary data $\mathring{m}_\cs,\mathring{n}_\cs$ which parametrize the shapes of boundary tetrahedra, and consider the branches $\a$ on which the boundary data can be achieved, i.e. $n_{\cs}^{(\a)}(\mathring{m})=\mathring{n}_{\cs}$, the last term in Eq.\Ref{S4d} on these branches $\a$ takes the same value, thus corresponds to an overall phase in $Z_{\sm_3}$ \cite{hanSUSY,3dblockHHKR}. This overall phase can again be removed in the asymptotics by project the partition function on coherent states in $m_\cs$ as in Eqs.\Ref{CohState} and \Ref{Q}.

The result Eq.\Ref{S4d} reproduces the earlier asymptotics result in \cite{hanSUSY}, which is obtained by semiclassically picking up by hand the branches $\a_{4d}$. Picking up $\a_{4d}$ semiclassically results in that the amplitude is only defined perturbatively via a semiclassical expansion. However here the result is achieved by a systematic quantization of simplicity constraint and imposing the constraint quantum mechanically to the amplitude. The resulting amplitude on simplicial complex is a non-perturbative definition, which hasn't been achieved in the earlier work. The above analysis shows that the branch $\a_{4d}$ stands out from the semiclassical approximation of the non-perturbative amplitude, which gives the correct semiclassical behavior.

\section{A Field-Theoretic Description of the Surface Defect}\label{field}

Recall that the insertion $\co_\cs$ in Eq.\Ref{insertion} projects Chern-Simons states on $\cs$ to the ground states $\psi_{\mathrm{sol}}$ of the ``Hamiltonian'' $H=\bfmu^2+\bfnu^2$. It has been mentioned that we can also introduce a squeezed version $H^{(w)}=w^{-2}\bfmu^2+w^2\bfnu^2$. The ground state $\psi^{w}_{\mathrm{sol}}$ of $H^{(w)}$ is a squeezed coherent state such that $\psi^{w=1}_{\mathrm{sol}}=\psi_{\mathrm{sol}}$. The squeezing parameter $w$ introduces an ambiguity to the model at each 4-holed sphere $\cs$ (in the following we equivalently understand $\cs$ as a sphere with 4 marked points, which are the intersections with the Wilson-lines, see Eq.\Ref{pathintegral}).

Here we would like to find a field-theoretic understanding of the surface defect $\co_\cs$, as well as the associated ambiguity.  The continuum counterparts of the conjugate variables $\mu,\nu$ are $\phi_1^i=\mathrm{Im}(A^i_1)$ and $\phi_2^i=\mathrm{Im}(A^i_2)$ (the coordinates on $\cs$ is chosen to be $x^{1,2}$). 
\be
\lt[\phi_1^i(x),\phi_2^j(x')\rt]=\frac{ik}{4\pi}\delta^{ij}\delta^{(3)}(x,x').
\ee
The Hamiltonian $H=\bfmu^2+\bfnu^2$ has the continuous conterpart $\int_\cs(\phi_1^i\phi_1^i+\phi_2^i\phi_2^i)$. However in order to make it coordinate-independent, we have to introduce a surface metric $h^{ab}$ ($a,b=1,2$), and write $\int_{\cs}\rmd^2x\,\sqrt{h}\, h^{ab}\phi_a^i\phi_b^i$. As a result the following operator plays the role as the projector $|\psi_{\mathrm{sol}}\rangle\langle \psi_{\mathrm{sol}}|$: 
\be
\co_\cs[A,\bA;h_{ab}]=\exp\lt[-\frac{k}{4\pi}\int_{\cs}\rmd^2x\,\sqrt{h}\, h^{ab}\phi_a^i\phi_b^i\rt],\quad a,b=1,2. \label{SurfaceOp}
\ee
The coupling constant has to be the same as the Chern-Simons level $k$. If we have chosen an independent coupling constant and scale it large, $\co_\cs$ would have been the same as inserting delta functions $\delta(\phi_1^i)\delta(\phi_2^i)$ in the path integral. However at the quantum level $\phi_1^i,\phi_2^i$ cannot be constrained to zero simultaneously by the uncertainty principle, since they are canonical conjugate variables. It relates to the zero-point energy of $H$. Letting the coupling constant the same as $k$ gives the sharpest projection. 

Here we find that the surface metric $h_{ab}$ is an analog or generalization of the above squeezing parameter $w$. Inserting $\co$ into the Chern-Simons theory breaks the topological invariance near the surface $\cs$, and makes the path integral explicitly depend on the metric $h_{ab}$ of each $\cs$. This metric dependence is the ambiguity of imposing simplicity constraint in the field-theoretic description.  

It is standard that the defect in Chern-Simons theory has certain metric-dependence, by breaking the topological invariance of Chern-Simons theory. An standard example is the Wilson line defect, whose metric dependence is reflected as the framing dependence. 

The defect might not depend on all the metric degrees of freedom, similar to the situation of Wilson lines. At the classical level, $\co_\cs$ is both conformal and reparametrization invariant on $\cs$. The metric dependence of $\co_\cs$ is essentially on the conformal equivalence classes of $h_\ab$. Two metrics from different classes are not related by conformal transformation and reparametrization. On a sphere with 4 marked points, one can always use conformal transformation move 3 marked points to 0,1 and $\infty$. The position of the last marked point on $S^2$, denoted by $\t$, labels the conformal equivalence classes of the metric. So the metric dependence of $\co_\cs$ is essentially on a single complex parameter in classical theory. It is interesting to understand whether this type of metric dependence is preserved at the quantum level, or how this property receives quantum corrections. The study of this point is postponed to the future research.

Explicitly we write the spinfoam amplitude on $\sm_4$ as a TQFT on 3-manifold $\Fx_3$ with both surface and line defects
\be
\sz_{\sm_3}=\int DA\, D\bA \  e^{-i\, CS [\mathfrak{X}_3| A,\bar{A}]}\prod_\cs \co_\cs[A,\bA;h_{\mu\nu}]\prod_l W_{(2j_l ,2\g j_{l})}[A,\bA]\label{pathintegral}
\ee
where the $\Slc$ Chern-Simons action reads
\be\label{eq:11}
CS\lt[\mathfrak{X}_3\big| A,\bar{A}\rt]=\frac{t}{8\pi}\int_{\Fx_3} \tr \lt(A\wedge dA+\frac{2}{3}A\wedge A\wedge A\rt)+\frac{\bar{t}}{8\pi}\int_{\Fx_3} \tr \lt(\bar{A}\wedge d\bar{A}+\frac{2}{3}\bar{A}\wedge \bar{A}\wedge \bar{A}\rt)
\ee
Here instead of defining the theory on $\sm_3$, we write the theory on the ambient space $\Fx_3$ and introduce a Wilson loop operator $W_{(j_l ,\g j_{l})}[A,\bA]$ for each torus cusps. The Wilson loops are traces of holonomies in the unitary representation $(2j_l ,2\g j_{l})$ of $\Slc$, where $\g=s/k$ is the Barbero-Immirzi parameter. In case $\sm_4$ has a boundary, Wilson line operators adjoint at vertices has to be introduces in $\Fx_3$ corresponding to the annuli cusps adjoint at 4-holed spheres as the boundary of $\sm_3$. For the simplicity of the following discussion, we focus on the case that $\sm_4$ has no boundary, so that $W_{(j_l ,\g j_{l})}[A,\bA]$ are all Wilson loops.

Indeed, inserting Wilson loops in TQFT on $\Fx_3$ is equivalent to TQFT on the complement $\sm_3=\Fx_3\setminus\{l\}$. It is standard that the Wilson loop operator has a path integral expression \cite{Dimofte}
\be
\prod_\ell W_{(2j_l ,2\g j_{l})}[A,\bA]=\int DY\,D\bar{Y}\,e^{\sum_{l}\frac{i}{2}\int_{\ell}\tr\lt[(\nu_l+\kappa_l)Y^{-1}(d+A)Y+(\nu_l-\kappa_l)\bar{Y}^{-1}(d+\bar{A})\bar{Y}\rt]},
\ee
where $\nu,\kappa$ relate to the representation labels $\nu_l=-\gamma j_l \sig_3,\ \kappa_l=i j_l \sig_3$ ($\sig_3$ is the 3rd Pauli matrix). $Y: \times_ll\to\Slc$ is a group-valued field. In the tubular neighborhood $N(l)$ of each Wilson loop, the Chern-Simons action can be written as
\be
\frac{t}{8\pi}\int_{N(l)}\tr\lt(A_\perp\wedge \rmd A_{\perp}\rt)+\frac{t}{4\pi}\int_{N(\ell)}\tr\lt(F_\perp\wedge A_{t}\rt)+\mathrm{c.c.}\label{CSN}
\ee
where $A_t,A_\perp$ are the components of $A$ along and perpendicular to $l$. $F_\perp=\rmd A_\perp+A_\perp\wedge A_\perp$ is the curvature. The above Chern-Simons action on $N(l)$ is coupled with the path integral of Wilson loop. The coupled action is linear to $A_t,\bA_t$, while other ingredients in $Z_{\sm_3}$ doesn't depend on $A_t,\bA_t$. $A_t,\bA_t$ can be integrated to get 2 delta functions constraining $F_\perp$ and $\bar{F}_\perp$ \cite{Dimofte}:
\be
\frac{t}{4\pi}F^T_\perp&=&\frac{1}{2}\sum_l Y\lt(\nu_l+\kappa_l\rt)Y^{-1}\delta_l^{(2)}(x)\rmd x_1\wedge \rmd x_2\nonumber\\
\frac{\bar{t}}{4\pi}\bar{F}^T_\perp&=&\frac{1}{2}\sum_l\bar{Y}\lt(\nu_l-\kappa_l\rt)\bar{Y}^{-1}\delta_l^{(2)}(x)\rmd x_1\wedge \rmd x_2.
\ee 
We have chosen a local coordinate $(x_1,x_2)$ on $D$ so that the Wilson line goes through the origin. The constraints imply the eigenvalue of median holonomy on each $T^2$ to be 
\be
\l_l=\exp\lt[\frac{2\pi i}{k}j_l\rt]
\ee
which is the boundary condition imposed to the theory on the complement $\sm_3=\Fx_3\setminus\{l\}$. $\l_l$ is the same as $\l_{T^2}$ in the last section. Equivalently $\sz_{\sm_3}$ can be written as a TQFT on $\sm_3$ with surface defects and the above boundary condition
\be
\sz_{\sm_3}=\int_{\l_l,\bar{\l}_{l}} DA\, D\bA\ e^{-i\, CS [\sm_3| A,\bar{A}]}\prod_\cs \co_\cs[A,\bA;h_{\mu\nu}].
\ee
The above is the field-theoretic version of the wave function Eq.\Ref{Zsm} defined in the previous sections. 

As $k,s\to\infty$ and keeping $\l_\ell$ fixed, the leading contribution of $Z_{\sm_3}$ comes from the solutions of critical equations $\delta S=\mathrm{Re} S=0$ when the path integral is written as $\int e^S$. $\mathrm{Re} S=0$ implies $\phi_a=0$ on each interface $\cs$, i.e. the connection reduces to SU(2) on $\cs$. At the solution of $\mathrm{Re} S=0$, the equation of motion $\delta S=0$ is simply the same as the Chern-Simons theory without surface defect, i.e. the connection is flat on $\sm_3$
\be
F=\bar{F}=0\ \ \text{on}\ \ \sm_3, \label{F=0}
\ee
and satisfies the boundary condition. It has been shown in the last section that all the flat connections satisfying the critical equations correspond to the simplicial geometries on $\sm_4$, although some flat connections may not give a uniform orientation $\sgn(V_4)$ on $\sm_4$.

As the semiclassical limit $k,s\to \infty$, the leading contribution of each critical point is given by evaluating the action at the critical point. In Eq.\Ref{pathintegral}, $\co_\cs=1$ at each critical point. The Chern-Simons action and the Wilson-loop action evaluated at a flat connection gives \cite{analcs,kirk}
\be
-\frac{t}{2\pi}\int_{c\subset\cl_\a}\ln\t_l\frac{\rmd\l_l}{\l_{l}}-\frac{\bar{t}}{2\pi}\int_{c\subset\cl_\a}\ln\bar{\t}_l\frac{\rmd\bar{\l}_l}{\bar{\l}_{l}}\label{AK}
\ee
where the integration is along a contour $c$ in the Lagrangian submanifold $\cl\simeq\cm_{flat}(\sm_3,\Slc)$. $\a$ labels the branch of $\cl$ where the flat connection locates. $\l_l,\t_l$ is the eigenvalues of meridian and longitude holonomies on the $T^2$ boundary. As a result, the contribution of a critical point gives the same result as Eqs.\Ref{S4d0} and \Ref{S4d} up to an overall constant (removing the boundary terms in Eqs.\Ref{S4d0} and \Ref{S4d}). The leading contribution of the flat connection $\a_{4d}$ gives the Regge action with cosmological constant on $\sm_4$ 
\be
\sz_{\sm_4}\sim \exp\frac{i}{\ell_P^2}\lt(\sum_{\ff}\mathbf{a}(\ff)\,\eps(\ff)-\L \sum_{\Fs}\lt|V_4(\Fs)\rt|\rt).
\ee

\section{Surface Degree of Freedom}\label{SDF}

The surface defect introduced in Eq.\Ref{SurfaceOp} explicitly breaks the $\Slc$ gauge invariance into SU(2) on the surface $\cs$. Then from the field theory point of view, the gauge degree of freedom becomes the physically propagating degree of freedom on $\cs$, similar to the case of 2d Wess-Zumino-Witten model as the boundary field theory of Chern-Simons theory in 3d bulk. In other words, introducing additional degree of freedom on $\cs$ recovers the $\Slc$ gauge invariance on $\cs$. 

We consider the infinitesimal gauge transformation of $\Slc$ connection, which turns out to be sufficient for the present purpose 
\be
\delta_\xi A_\mu^i= D_\mu\xi^i=\partial_\mu\xi^i+\eps^{ijk}A_\mu^j\,\xi^k,\quad \delta_{{\xi}} \bar{A}_\mu^i= \bar{D}_\mu\bar{\eps}^i=\partial_\mu\bar{\xi}^i+\eps^{ijk}\bar{A}_\mu^j\,\bar{\xi}^k
\ee
We consider the background field $(A,\bA)$ being a critical point of the path integral, which satisfies $\phi_a^i=0$ on $\cs$. So we have $A_a^i$ is an SU(2) connection, and $D_a=\bar{D}_a$ is an SU(2) covariant derivative on $\cs$ with respect to the background field. Therefore the gauge transformation of $\phi_a^i$ is
\be
\delta_\xi\phi_a^i=\frac{1}{2 i}D_a \lt(\xi^i-\bar{\xi}^i\rt)\equiv D_a \varphi^i
\ee
where $\varphi^i=\frac{1}{2 i}\lt(\xi^i-\bar{\xi}^i\rt)$ is a scalar in adjoint representation of SU(2). 

The infinitesimal gauge transformation of $\co_\cs$ in Eq.\Ref{SurfaceOp} gives 
\be
\delta_{\varphi}\lt(-\frac{k}{4\pi}\int_{\cs}\rmd^2x\,\sqrt{h}\, h^{ab}\phi_a^i\phi_b^i\rt)=-\frac{k}{4\pi}\int_{\cs}\rmd^2x\,\sqrt{h}\, h^{ab}D_a \varphi^iD_b \varphi^i + o(\varphi^3) \label{DphiDphi}
\ee
at the critical background field with $\phi_a^i=0$ on $\cs$. If we add it to the exponent of Eq.\Ref{SurfaceOp} and redefine $\co_\cs$ by
\be
\co_\cs\lt[A,\bA,h_{ab}\rt]:=\int D\varphi^i\ \exp\lt[-\frac{k}{4\pi}\int_{\cs}\rmd^2x\,\sqrt{h}\, h^{ab}\phi_a^i\phi_b^i-\frac{k}{4\pi}\delta_{\varphi}\lt(\int_{\cs}\rmd^2x\,\sqrt{h}\, h^{ab}\phi_a^i\phi_b^i\rt)\rt].
\ee
Now $\co_\cs\lt[A,\bA,h_{ab}\rt]$ is invariant under $\Slc$ gauge transformation (any gauge transformation can be compensated by a shift of gauge parameter $\varphi$). Expanding the term $\delta_{\varphi}\lt(\int_{\cs}\rmd^2x\,\sqrt{h}\, h^{ab}\phi_a^i\phi_b^i\rt)$ at the critical background field gives Eq.\Ref{DphiDphi} as the leading order for small $\varphi$. The additional term in $\varphi$ looks like a (gauged) linear sigma-model on $\cs$.

When we insert the above complete operator $\co_\cs\lt[A,\bA,h_{ab}\rt]$ into Eq.\Ref{pathintegral}, the additional degree of freedom $\varphi$ on $\cs$ doesn't modify our previous semiclassical analysis. The Chern-Simons connections $A$ that we are interested in are nontrivial on all 4-holed spheres $\cs$. Turning on a nontrivial background field $A_a\neq 0$ on $\cs$ makes $\varphi$ massive, whose mass term is given by 
\be
\eps^{ijk}\eps^{ilm}h^{ab}A_a^jA_b^l\varphi^k\varphi^m
=h^{ab}\lt(\delta^{jl}\delta^{km}-\delta^{jm}\delta^{kl}\rt)A_a^jA_b^l\varphi^k\varphi^m
=\lt[h^{ab}\lt(\delta^{km}A_a^lA_b^l-A_a^mA_b^k\rt)\rt]\varphi^k\varphi^m
\ee
One can diagonalize the mass matrix in the square bracket by orthogonal transformation $N$, i.e. \\$N^{-1}\lt[\tr(A^T h A)\mathbf{1}-A^T hA\rt]N=\tr(A^T h A)\mathbf{1}-\mathrm{diag}(x_1,x_2,x_3)$, where $x_{1,2,3}\geq0$. and $\tr(A^T h A)=x_1+x_2+x_3>0$\footnote{$A^T h A$ is a positive semi-definite matrix when $A\neq0$. $x_i$ may vanishes since $A_a^i$ may not be a nondegenerate matrix. But if all $x_{1,2,3}=0$, it would lead to $(AN)^Th(AN)=0$, which implies $AN=0$ and $A=0$, since $N$ is invertible.}. So the eigenvalues of the mass matrix are all positive. $\varphi$ being massive motivates us to integrate out $\varphi$, which at the semiclassical level projects to the ground state $\varphi=0$. 

The surface defect $\co_\cs$ modifies the equations of motion by adding a singular term
\be
F(A)=\delta(t)\,\rmd t\wedge J\lt(h_{ab},\phi_a^i,\varphi^i\rt)
\ee 
where $t$ is the coordinate transverse to $\cs$, and the location of $\cs$ corresponds to $t=0$. The critical equation $\phi_a^i=0$ and $\varphi=0$ on $\cs$ implies $J\lt(h_{ab},\phi_a^i,\varphi^i\rt)=0$ on the right-hand side of the equation of motion. Thus the equation of motion reduces to the flatness Eq.\Ref{F=0}. So we conclude that all the critical flat connections on $\sm_3$ studied in the last section are still critical, even when we take into account the additional degree of freedom $\varphi$ on the surface defect.

\section{Conclusion and Outlook}

In this paper we study the quantization and implementation of LQG simplicity constraint in spinfoam model in the presence of cosmological constaint. Spinfoam amplitudes with cosmological constant are formulated as complex Chern-Simons theories on certain class of 3-manifolds. Implementation of quantum simplicity constraint results in surface defects in the Chern-Simons theory. These surface defects guarantee the amplitude have the correct semiclassical limit, which reproduces the Einstein-Regge action with cosmological constant on 4d simplicial complex.

This work relates LQG simplicity constraint to surface defects in Chern-Simons theory. Although line defects have been widely studied in Chern-Simons theory, surface defects (or domain-walls) are however not sufficiently studied in the literature. The surface defect appearing here has not been studied before. We have done some preliminary investigations of the surface defect by studying the propagating physical degrees of freedom on the defect surface. We show that at the linearized level, the propagating field behaves as a 2d sigma model gauged by Chern-Simons connection.

The formalism in this paper makes it possible to define rigorously the spinfoam amplitude with cosmological constant. The present definition of the amplitude either uses the infinite dimensional path integral \cite{HHKR} or uses a semiclassical expansion \cite{HHKRshort}. However it is known that the Chern-Simons partition function $Z_{S^3\setminus\G_5}$ can be expressed as a finite dimensional integral \cite{hanSUSY}. Now the spinfoam amplitude is constructed by projecting $Z_{S^3\setminus\G_5}$ onto the solution of simplicity constraint by Eq.\Ref{Zint}, which is also a well defined operation. So the entire spinfoam amplitude can be written as a finite dimensional integral, whose finiteness is ready to be studied. The research on this aspect is currently undergoing.

Further studies of the proposed surface defect is also postponed to future research: It is interesting to understand the dynamics of the sigma model propagating on the defect, including its interaction with Chern-Simons theory. The metric dependence of the surface defect might be understood more detailedly in the future. The defect may not depend on all the metric degrees of freedom (like the situation of Wilson lines). Classically the surface defect only depends on the complex structure of the 4-holed sphere, since the defect is both reparametrization and conformal invariant classically. Then it is interesting to see whether this type of metric dependence is preserved at the quantum level, or how this property receives quantum corrections.

%%%%%%%%%%%%%%%%%%%%%%%%%%%%%%%%%%%%%%%%%%%%%%%%%%%%%%%%%%%%%%%%%%%%%%%%%%%%%%%%%%%%%%%%%%%%%%%%%%%%%%%%%%%%%%%%%%%%%%%%%%%%%%%

%%%%%%%%%%%%%%%%%%%%%%%%%%%%%%%%%%%%%%%%%%%%%%%%%%%%%%%%%%%%%%%%%%%%%%%%%%%%%%%%%%%%%%%%%%%%%%%%%%%%%%%%%%%%%%%%%%%%%%%%%%%%%%%

%%%%%%%%%%%%%%%%%%%%%%%%%%%%%%%%%%%%%%%%%%%%%%%%%%%%%%%%%%%%%%%%%%%%%%%%%%%%%%%%%%%%%%%%%%%%%%%%%%%%%%%%%%%%%%%%%%%%%%%%%%%%%%%

%%%%%%%%%%%%%%%%%%%%%%%%%%%%%%%%%%%%%%%%%%%%%%%%%%%%%%%%%%%%%%%%%%%%%%%%%%%%%%%%%%%%%%%%%%%%%%%%%%%%%%%%%%%%%%%%%%%%%%%%%%%%%%%

\section*{Acknowledgements}

MH acknowledges Xin Gao, Du Pei, Jian Qiu, and Junya Yagi for useful discussions. MH also acknowledges Ling-Yan Hung, Yidun Wan at Fudan University, Yong-Shi Wu at the University of Utah for invitations, hospitality during his visits, and inspiring discussions. This work receives supports from the US National Science Foundation through grant PHY-1602867, and Start-up Grant at Florida Atlantic University, USA.

%\appendix

%\nocite{*}

%\bibliographystyle{jhep}
%\bibliographystyle{hep}
%\bibliography{muxin}

\providecommand{\href}[2]{#2}\begingroup\raggedright\endgroup

\end{document}